\documentstyle[multicol,aps,epsf]{revtex}

\begin{document}
\title{Gradient Clogging in Depth Filtration}
\author{S.~Datta and S.~Redner}

\address{Center for Polymer Studies and Department of Physics, Boston
University, Boston, MA, 02215}
\maketitle
\begin{abstract}

We investigate clogging in depth filtration, in which a dirty fluid is
``cleaned'' by the trapping of dirt particles within the pore space
during flow through a porous medium.  This leads to a gradient
percolation process which exhibits a power law distribution for the
density of trapped particles at downstream distance $x$ from the input.
To achieve a non-pathological clogging (percolation) threshold, the
system length $L$ should scale no faster than a power of $\ln w$, where
$w$ is the width.  Non-trivial behavior for the permeability arises only
in this extreme anisotropic geometry.

\end{abstract}
\begin{multicols}{2}

Depth filtration is a mechanism for separating suspended particles from
a carrier fluid by passing it through a porous
medium\cite{rev1,rev2,rev3}.  The medium promotes efficient filtering
both by increasing the area available for deposition of suspended
particles, as well as the exposure time of the suspension to the
absorbing interfaces.  This mechanism is therefore widely used in a
variety of biological, chemical, and engineering separation
processes\cite{rev1}.  Depth filtration also raises basic issues in
porous media transport, as the medium becomes progressively constricted
by particle trapping events so that a steady state is not achieved.
This feedback between flow and structure governs the essential
properties of filtration.  Each pore blockage results in a small
reduction of the permeability of the medium and ultimately a clogging
(percolation) threshold is reached where the filter permeability
vanishes.  Previous studies indicated that the permeability vanishes as
a power law near the threshold, with an exponent different from that of
classical percolation\cite{sahimi,mft}.  From a practical perspective,
filter performance is degraded by particle trapping, so that
understanding this trapping rate is of paramount importance.

In this Letter, we formulate a geometrical description for depth
filtration which provides intuition for the clogging process and leads
to phenomenology outside the realm of classical percolation.  For
overlapping distributions of pore and particle radii, the trapped
particle density distribution varies as a power law in longitudinal
coordinate.  Such a distribution should be readily observable, for
example, when opaque particles are passed through a glass bead-pack
porous medium\cite{beads}.  This gradient leads to an unusual
percolation process where the percolation threshold (fraction of open
pores) is very close to unity and where the permeability does not have
power law behavior.

Of the many microscopic interactions that underlie filtration, we focus
on size exclusion\cite{sahimi,lk}, where a particle of radius $r_{\rm
particle}$ is trapped within the first pore encountered whose radius
satisfies $r_{\rm pore}<r_{\rm particle}$.  This size exclusion is the
dominant effect in processes such as gel permeation in porous media and
liquid chromatography.  While other influences, such as Van der Waals
forces between particles and pore surfaces, hydrodynamic and
electrostatic interactions, {\it etc.}, may be important, their faithful
modeling is complex\cite{rev1} and makes it difficult to identify the
governing mechanism for a given macroscopic property.  Our approach is
to retain size exclusion as the only trapping mechanism in a geometric
modeling of filtration and develop physical intuition for clogging from
this idealized description.

\begin{figure}
\narrowtext
\epsfxsize=2.5in\epsfysize=0.7in
\hskip 0.3in\epsfbox{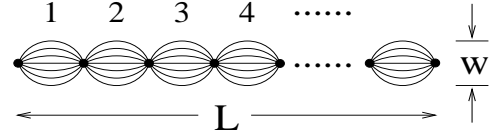}
\vskip 0.15in
\caption{The bubble model.  Each line represents fluid carrying pore. 
\label{fig1}}
\end{figure}

We first introduce the quasi one-dimensional ``bubble'' model to
describe filtration.  This system consists of $L$ links in series, in
which each link is a parallel bundle of $w$ bonds and each bond
represents a pore (Fig.~1).  This model can be viewed as a square
lattice in which all perpendicular bonds are ``shorted''.  This system
was introduced to account for the breaking of fibers\cite{fiber} and to
determine extremal voltages associated with breakdown in
resistor-fuse networks\cite{bubble}.  An appealing feature of this model
in the context of percolation is that it exhibits finite-dimensional
behavior when $L$ scales as $e^w$.  Namely, if each bond is randomly
occupied with probability $p$, a percolation threshold at a value of
$p_c$ strictly between 0 and 1 arises, and associated critical exponents
can be easily computed\cite{bubble}.

To adapt this model to filtration, we posit that each bond has a radius
$r$ drawn from a specified distribution, with volumetric flow rate
proportional to $r^4\,\nabla p$ (Poiseuille flow), where $\nabla p$ is
the pressure gradient in the bond.  Dynamically neutral suspended
particles move through the medium at a rate governed by this local flow.
We assume perfect mixing at each node, in which a suspended particle has
a flow induced probability proportional to $r_i^4$ to enter an unblocked
bond of radius $r_i$ in the next downstream bundle.  The particles,
whose radii are also drawn from a distribution, are injected singly and
tracked until each is trapped or escapes the system.  Upon capture, the
particle is defined to block the bond completely so that there is no
further fluid flow in this bond.  After each blockage event, the new
flow field is computed to determine the trajectory of the next suspended
particle.

This system exhibits three regimes of behavior.  For pores typically
smaller than particles (subcritical), the particles get trapped almost
immediately and rapid clogging ensues.  Conversely, for pores typically
larger than particles (supercritical), a steady state is eventually
reached for a finite length system, in which the smallest pores are
blocked and the suspension flows freely through the remaining
unblockable pores.  These cases can be viewed as corresponding to poor
filter performance.  At the boundary between these regimes is the
critical case, where the particle and pore radius distributions overlap
substantially.  Here, particle trapping is gradual, with considerable
penetration of the medium before clogging is reached.  This may be
viewed as efficient filtration because of the large number of particles
filtered before clogging and the relatively long filter lifetime.  Thus
both from practical and theoretical perspectives, the critical case is
the most interesting.

For simplicity and concreteness, consider a uniform distribution of both
particle and bond radii in the range $[a,b]$.  More general continuous
distributions can be straightforwardly treated, but little new
qualitative insight emerges.  Let us first determine the spatial
distribution of trapped particles during filtration.  The gradient
nature of the trapping process implies that the number of blocked bonds
in downstream bubbles remains small, even at the percolation threshold
(see Fig.~2).  We therefore employ an ``unperturbed'' approximation in
which the initial bond radius distribution is continued to be used
during the clogging process.  Within this approximation and assuming
Poiseuille flow, the probability that a particle of radius $r$ gets
trapped in a bubble is
\begin{equation}
\label{trap}
P_< = {\int_a^r r'^4 dr' \over \int_a^b r'^4 dr'} = {r^5 - a^5 \over b^5 - a^5},
\end{equation}
for large $w$.  (Exact calculation over all configurations of bond radii
shows that the above large-$w$ form is asymptotically correct for $w\geq
5$.)~ Consequently, the probability that a particle gets trapped in the
$n^{\rm th}$ bubble is
\begin{equation}
\label{pn}
P_n = (1 - P_<)^{n-1} P_<.
\end{equation}
Averaging over the distribution of particle radii gives, 
\begin{eqnarray}
\label{pnav}
\langle P_n \rangle &=& \int_a^b \left(1-{r^5-a^5 \over b^5-a^5}\right)^{n-1}
{r^5-a^5 \over b^5-a^5} \,{dr \over b-a},\nonumber\\
            &=& {1\over 5} {(b^5-a^5)\over (b-a)} \int_0^1 {v
(1-v)^{n-1} dv \over [v(b^5-a^5) + a^5]^{4 \over 5}} ,
\end{eqnarray}
where $v={r^5-a^5 \over b^5-a^5}$.  

Depending on the lower cutoff $a$, there are two different asymptotic
behaviors for this trapping probability.  If $a=0$, the integral reduces
to the beta function \cite{as}
\begin{equation}
\label{pnf}
\langle P_n \rangle =  {\Gamma({6 \over 5}) \Gamma(n)
\over 5\,\Gamma(n + {6 \over 5})},
\end{equation}
where $\Gamma(n)$ the gamma function, and for large $n$, $\langle
P_n\rangle\sim 0.1836\ldots n^{-6/5}$.  Conversely, if $a \ne 0$, then the
asymptotic behavior of the integral in Eq.~(\ref{pnav}), which arises
from the contribution near the lower limit, can be written approximately
as
\begin{equation}
\label{pnappr}
\langle P_n \rangle \approx   {b^5-a^5\over{5a^4(b-a)}} \int_0^1 v
(1-v)^{n-1} \,dv \propto n^{-2}.
\end{equation}

\begin{figure}
\narrowtext
\epsfxsize=1.9in\epsfysize=2.0in
\hskip -0.0in\epsfbox{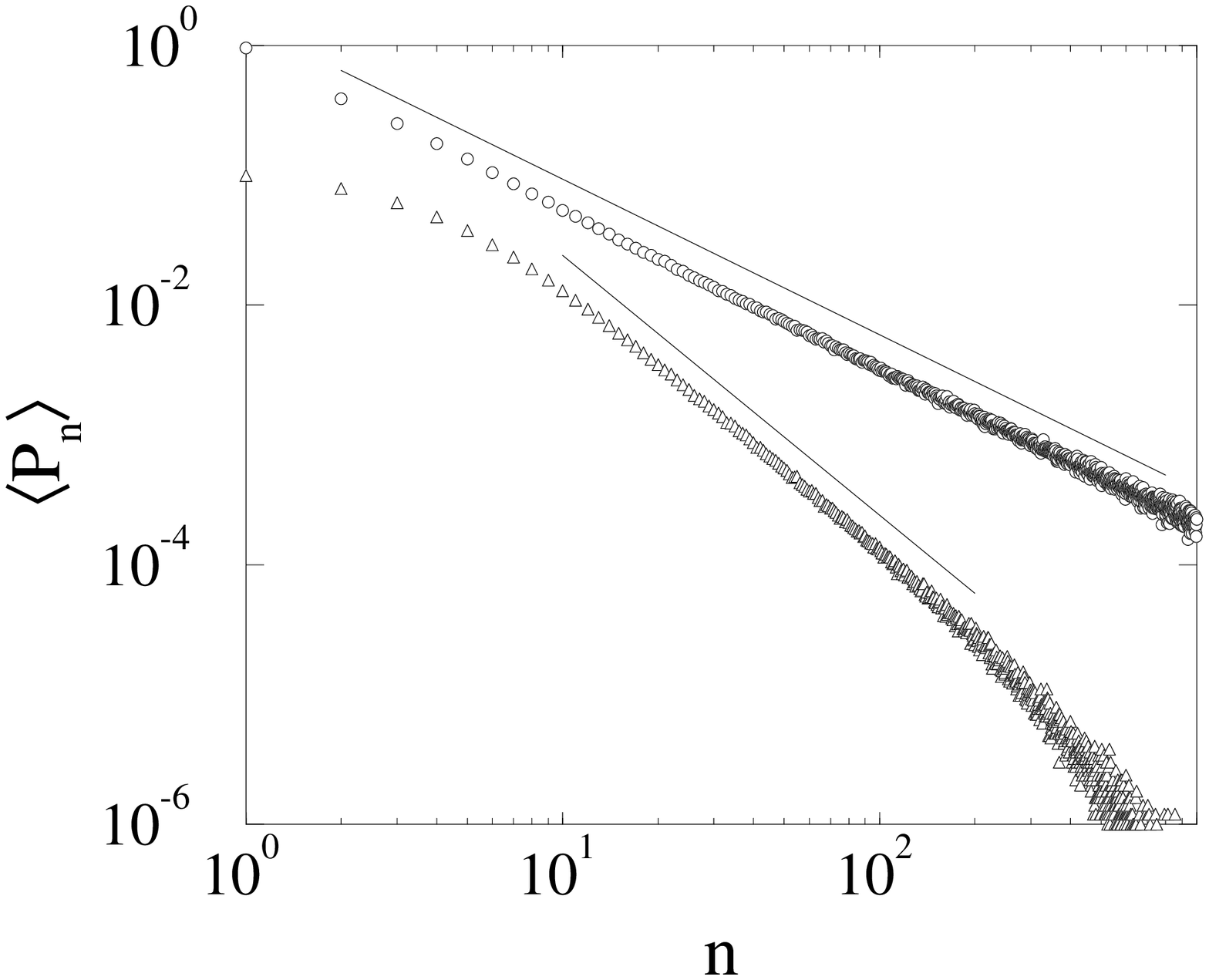}
\epsfxsize=1.9in\epsfysize=2.0in 
\hskip -0.47in\epsfbox{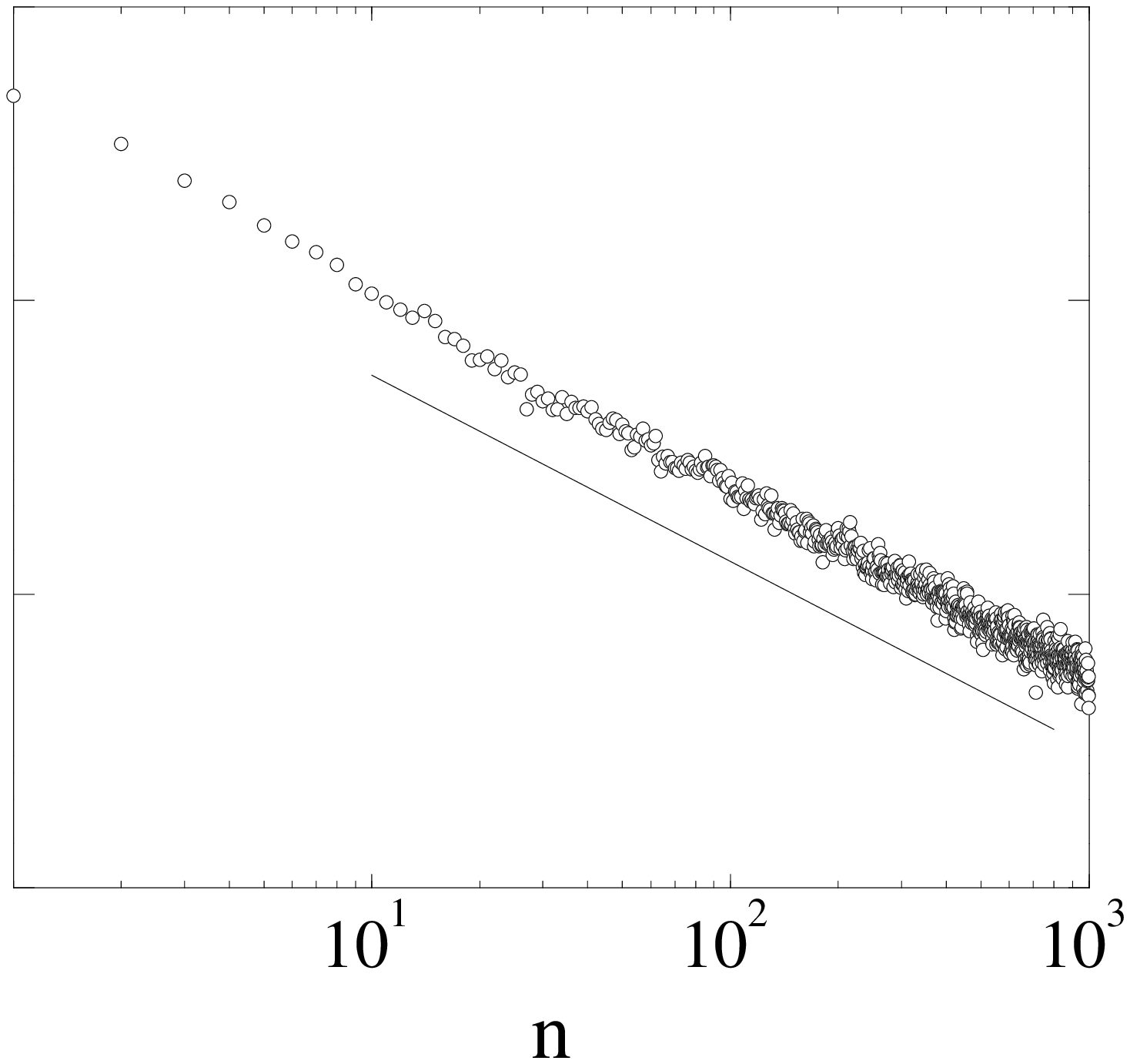} 
\vskip 0.15in
\caption{(a)  Trapping probability $\langle P_n\rangle$ versus $n$ 
at the percolation threshold for a bubble model of width $w=50$ for
coincident and uniform distributions of particle and bond radii.  Shown
are data, based on $10^4$ configurations, for $(a,b)=(0,1)$ ($\circ$)
and $(a,b)=(0.7,1.0)$ ($\Delta$), with the latter divided by 10 for
visualization.  The straight lines have slopes $-6/5$ and $-2$
respectively.  (b) Trapping probability, based on injecting $4\times
10^6$ particles into an unperturbed $500\times 1000$ square lattice
whose axes are oriented at $45^\circ$ with respect to the average flow,
for coincident, uniform particle and bond radius distributions on
$(0,1)$.  The straight line has slope $-1.27$.
\label{fig2}}
\end{figure}

From the denominator in the second line of Eq.~(\ref{pnav}), the
crossover between the $n^{-6/5}$ and $n^{-2}$ behaviors occurs when
$v(b^5-a^5) < a^5$, or equivalently, $n>n^{\ast}= (b/a)^5$.  While the
exponents $6/5$ and 2 are specific to the uniform radius distribution
and the flow-induced bond entrance probability, the existence of the
power law is generic and requires only the overlap of the bond and
particle radius distributions.  For example, for the Hertz distribution
of particle and bond radii, $p(r)=2re^{-r^2}$, the result corresponding
to Eq.~(\ref{pnf}) is $\langle P_n\rangle\propto n^{-4/3}$.

Qualitatively similar behavior for $\langle P_n\rangle$ occurs in
lattice networks.  In the spirit of our unperturbed approximation and to
obtain relatively extensive data, we focus on the spatial distribution
of the initially injected particle.  Later particles exhibit nearly the
same spatial distribution of trapping location, but much more time is
needed for computing this distribution, since the network permeability
must be recalculated after each trapping event.  For $n\agt 10$, the
best fit power law to the data is $\langle P_n\rangle\sim n^{-\mu}$,
with $\mu\approx 1.27$ (Fig.~2(b)).

In the supercritical regime (bonds larger than particles), $\langle
P_n\rangle$ exhibits near-critical behavior, except that some particles
can escape from the system.  Conversely, in the subcritical regime
(bonds smaller than particles), Eq.~(\ref{pnav}) gives $\langle
P_n\rangle \propto \exp[-n(a^5-A^5)/(B^5-A^5)]$, where $(a,b)$ and
$(A,B)$ are, respectively, the ranges of the particle and bond radius
distributions.  As $a-A\to 0$, the decay length $(B^5-A^5)/(a^5-A^5)$
diverges and power law behavior of $\langle P_n\rangle$ is recovered.
It is in this sense that coincident bond and particle radius
distributions corresponds to a critical phenomenon.

Let us now determine the number of particles that need to be injected to
reach the clogging (percolation) threshold.  For the bubble model, this
means that all bonds in a single bubble are blocked.  Since $\langle
P_n\rangle$ monotonically decreases in $n$, the probability that all $w$
bonds are blocked in the $n^{\rm th}$ bubble is non-zero only for small
$n$.  (Numerically, for $w=100$, for example, the probability of
blocking in bubble $n$ is approximately $78.2\%$, $15.8\%$, $4.44\%$,
and $1.21\%$ for $n=1$, 2, 3, and $4$, respectively.)~ In the following,
we therefore work within the approximation that it is only the first
bubble that clogs, and that both the particle and bond radius
distributions are uniform on $[0,1]$.

We first compute the number of particles that need to be injected into a
bond of radius $0<r<1$ before it becomes blocked\cite{extreme}.  For $N$
particles, the probability that all have their radii in the range
$[0,r]$ is simply $r^N$.  This can be re-interpreted as the probability
that the maximum radius among $N$ particles lies between 0 and $r$.
That is $r^N=\int_0^rP_N(r')\,dr'$, with $P_N(r)$ the probability
density that the maximum radius equals $r$.  Consequently,
$P_N(r)=Nr^{N-1}$, and the average radius of this largest particle is
\begin{equation}
\label{rmax}
\langle r\rangle_N =\int_0^1r\,P_N(r)\,dr = {N\over {N+1}}.
\end{equation}
Inverting this relation shows that of the order of $(1-r)^{-1}$
particles need to be injected before a particle of sufficiently large
radius enters to block a bond of radius $r$.

Consider now a single bubble of $w\gg 1$ bonds.  The number of particles
needed to block bonds whose radii are in the range $[r,r+dr]$ is
$w{dr\over{1-r}}$.  Consequently, the total number of particles needed
to block the bubble is
\begin{equation}
\label{ncint}
N_c\approx \int_{1/w}^{1-1/w} w\,{dr\over{1-r}}.
\end{equation}
Here we again use Eq.~(\ref{rmax}) to determine that the largest and
smallest bond radii in the bubble are $r_{\rm max}\approx 1-1/w$ and
$r_{\rm min}\approx 1/w$, respectively.  The integral is dominated by
the behavior at the upper limit and gives
\begin{equation}
\label{nc}
N_c\propto w\ln w.
\end{equation}
Notice that a naive determination of the percolation threshold from
$N_c\langle P_1\rangle=w$ gives, using Eq.~(\ref{pnf}), $N_c= 6w$.  The
logarithmic factor in Eq.~(\ref{nc}) arises from the widest bonds for
which many particles need to be injected before blocking occurs.

\begin{figure}
\narrowtext
\epsfxsize=2.5in\epsfysize=1.8in
\hskip 0.3in\epsfbox{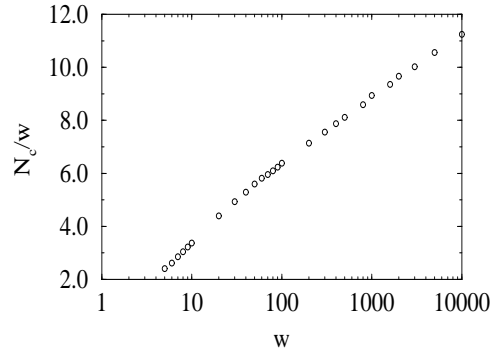}
\vskip 0.15in
\caption{Number of particles injected at percolation  (normalized by 
$w$) versus $\ln w$.  The number of configurations is $10^5$ for
$w<10^2$, $10^4$ for $10^2\leq w\leq 10^3$, and $10^3$ for $w>10^3$.
\label{fig3}}
\end{figure}

Simulations on the bubble model indicate that this logarithmic $w$
dependence for $N_c$ is independent of the precise form of the entrance
probability for a particular bond and similar details.  Thus if the
system size increases isotropically, the percolation threshold
$p_c=1-N_c/Lw$ approaches 1, with $Lw$ is the total number of bonds in
the system.  To obtain a threshold value less than unity requires
exponential anisotropy in which $L\sim\ln w$.  This result is
qualitatively robust with respect to different particle and bond radius
distributions.  For example, for the Hertz distribution, following
analogous computations to those just outlined gives $N_c\propto w (\ln
w)^2$.  For the square lattice, on the other hand, simulations indicate
that $N_c$ is linearly proportional to the system width.  This
corresponds to a percolation threshold which scales as $1-1/L$.  Thus
either $L$ should be constant, or an alternative relation between the
bond and particle radius distributions may be appropriate to define
criticality for a finite-dimensional network.

Finally, consider the behavior of the permeability during filtration,
for which the bubble model again provides a useful starting description.
Due to the series geometry of the bubble model, the permeability
$\kappa$ can be written as
\begin{equation}
\label{perm}
\kappa = [\sum_{n=1}^{L} ({1 \over k_{n}})]^{-1},
\end{equation}
where $k_n$ denotes the permeability of the $n^{\rm th}$ bubble.  As the
filter becomes constricted, the number of trapped particles in the
$n^{\rm th}$ bubble is proportion to $n^{-\mu}$ (Eq.~(\ref{pnf})), with
$\mu$ detail dependent.  Numerically, we find that the permeability of
the $n^{\rm th}$ bubble also scales as a power of the number of
unblocked bonds, $k_n\propto w-A/n^{\nu}$.  Here $\nu$ is also detail
dependent and $A$ is proportional to the total number of particles
trapped, with $A\to w$ corresponding to clogging (up to logarithmic
factors).  Thus the inverse permeability is
\begin{equation}
\label{perma}
\kappa^{-1} \sim \int_1^L{dn\over{w-A/n^\nu}}.
\end{equation}

This integral is approximately constant and gives $\kappa^{-1}\sim L/w$,
{\em except} close to clogging.  To estimate the integral in this limit,
note that for $n$ close to one, the integrand is dominated by the
divergence in the denominator, while for $n>(A/w)^{1/\nu}$ the second
term in the denominator can be neglected.  Splitting up the integral
according to this prescription gives
\begin{equation}
\label{permb}
\kappa^{-1} =
\int_1^{n^\ast} {dn\over{w-A/n^\nu}} + \int_{n^\ast}^L {dn\over w},
\end{equation}
with $n^\ast= (A/w)^{1/\nu}$.  The first integral may be estimated by
defining $v=wn^\nu/A$ and then treating the resulting slowly varying
factor of $v^{1/\nu}$ in the numerator as constant compared to the
divergent factor $1/(v-1)$.  We thereby obtain
\begin{equation}
\label{permc}
\kappa^{-1}\sim {L\over w}\left[1-\left({A\over w}\right)^{1/\nu}{1\over
L}\ln\left({w\over A}-1\right)\right],
\end{equation}
where the correction term in $\kappa^{-1}$ is manifestly positive near
the clogging threshold ($A\to w$ from above).  This crude estimate shows
that the permeability of an isotropic system ($L\propto w$) is
essentially unaffected by individual bond blocking events until one
bubble is nearly completely blocked, after which $\kappa$
discontinuously drops to zero.  If, however, $L\sim\ln w$, then $\kappa$
vanishes as a logarithm in $A-w$, where $A-w$ can be identified as the
distance to the percolation threshold.

In summary, a quasi-one-dimensional bubble model successfully describes
various geometrical aspects of depth filtration.  For coincident bond
and particle radius distributions, the number of particles trapped a
distance $n$ downstream asymptotically varies as $n^{-\mu}$, with $\mu$
dependent on details of these distributions.  The percolation threshold
can be determined from extreme value considerations, and within the
bubble model, the length must scale as the logarithm of the system width
to have a percolation threshold strictly less than unity.  For such a
geometry, the permeability exhibits logarithmic dependence on $(p-p_c)$
over a restricted range.  These results may help explain previous
simulations on relatively small systems\cite{sahimi}, where a threshold
value close to unity and a permeability which varied rapidly in
concentration near the threshold was observed.

These unusual results are a consequence of the gradient nature of the
particle trapping process which predominantly affects the upstream
portion of the network.  This gradient aspect also has ramifications for
efficient filter design.  A system whose width is much greater than its
length is needed to give a percolation threshold less than unity.  This
geometry has the desirable feature that a finite fraction of the medium
actually traps particles.  On the other hand, the radius-average
probability that a particle escapes a system of length $L$ vanishes as
$L^{1-\mu}$ for $\langle P_n\rangle\sim n^{-\mu}$.  Thus a small escape
rate and a percolation threshold strictly less than unity, which
ostensibly are the desired properties of a depth filter, cannot
simultaneously be satisfied in a spatially homogeneous medium.  This
suggests that an optimal filter should have a longitudinal varying local
permeability which effectively cancels the gradient nature of particle
trapping process.

We thank J. Koplik and P. L. Krapivsky for helpful discussions and a
critical reading of the manuscript.  We also gratefully acknowledge NSF
grant number DMR-9632059 for financial support.

\end{multicols} 


\begin{thebibliography}{99}
\medskip

\bibitem{rev1} C. Tien, {\sl Granular Filtration of Aerosols and Hydrosols}
(Butterworths, Boston, 1989); see also C. Tien and A. C. Payatakes,
AICHE J. {\bf 25}, 737 (1979).

\bibitem{rev2} J. Dodds, G. Baluais and D. Leclerc, in {\sl Disorder and
Mixing}, eds.\ E. Guyon, J.-P. Nadal and Y. Pomeau, (Kluwer, Dordrecht,
1988); D. Houi, in {\sl Hydrodynamics of Dispersed Media}, ed.  J. P.
Hulin, A. M. Cazabat, E. Guyon and F. Carmona (Elsevier, Amsterdam,
1990); J. H. D. Hampton, S. B. Savage and R. A. L. Drew,
Chem. Engr.\ Sci.\ {\bf 48}, 1601 (1993).

\bibitem{rev3} M. Sahimi, Rev.\ Mod.\ Phys. {\bf 65}, 1393 (1993).

\bibitem{sahimi} A. O. Indakm and M. Sahimi, Phys.\ Rev.\ A {\bf 36},
5304 (1987); Phys.\ Rev.\ Lett. {\bf 66}, 1169 (1991).

\bibitem{mft} M. M. Sharma and Y. C. Yortos, AIChE J. {\bf 33}, 1637,
1644, 1654 (1987).

\bibitem{beads} C. Ghidaglia, E. Guazelli and L. Oger, J. Phys.\ D {\bf
24}, 2111 (1991); C. Ghidaglia, E. Guazelli, L. de Arcangelis and L.
Oger, to be published; C. Ghidaglia, Th\`{e}se de Doctorat de
l'Universit\'{e} Paris VI (1994).

\bibitem{lk} J. Lee and J. Koplik, Phys.\ Rev.\ E {\bf 54}, 4011 (1996).
(1997).

\bibitem{fiber} H. E. Daniels, Proc.\ Roy.\ Soc., Ser.\ A  {\bf
183}, 404 (1945); D.~G. Harlow and S. L. Phoenix,  J.\ Comput.\
Mater.\ {\bf 12}, 195 (1978); S.~L. Phoenix and R.~L. Smith, J. Appl.\
Mech.\ {\bf 103} 75, (1981); D. Sornette and S. Redner, J. Phys.\ A {\bf
22}, L619 (1989); P.~L. Leath and P.~M. Duxbury, Phys.\ Rev.\ B {\bf 49},
14905 (1994).

\bibitem{bubble} B. Kahng, G. G. Batrouni, and S. Redner, J. Phys.\ A
{\bf 20}, L827 (1987).

\bibitem{as} M.~Abramowitz and I.~A.~Stegun, {\sl Handbook of  Mathematical
Functions} (Dover, New York, 1965).

\bibitem{extreme} This is a basic exercise in extreme value statistics.
See {\it e.g.}, J. Galambos, {\sl The Asymptotic Theory of Extreme Order
Statistics}, (J. Wiley \& Sons, New York, 1978).



\end{thebibliography}
\end{document}